\documentclass[aps,prl,twocolumn,amsfonts,nobibnotes,superscriptaddress,showpacs]{revtex4-1}

\usepackage{dcolumn}
\usepackage{amsmath}
\usepackage{amssymb}
\usepackage{graphicx}
\usepackage{bm}
\usepackage{color}

\begin{document}

\title{Giant Exciton Mott Density in Anatase TiO$_2$}

\vspace{2cm}

\author{Edoardo Baldini$^*$}
\affiliation{Laboratory of Ultrafast Spectroscopy, ISIC and Lausanne Centre for Ultrafast Science (LACUS), \'Ecole Polytechnique F\'{e}d\'{e}rale de Lausanne, CH-1015 Lausanne, Switzerland}

\author{Tania Palmieri}
\affiliation{Laboratory of Ultrafast Spectroscopy, ISIC and Lausanne Centre for Ultrafast Science (LACUS), \'Ecole Polytechnique F\'{e}d\'{e}rale de Lausanne, CH-1015 Lausanne, Switzerland}

\author{Adriel Dominguez}
\affiliation{Bremen Center for Computational Material Science (BCCMS), Bremen, Germany}
\affiliation{Shenzhen JL Computational Science and Applied Research Institute (CSAR), Shenzhen, China}
\affiliation{Beijing Computational Research Center (CSRC), Beijing, China}

\author{Angel Rubio}
\affiliation{Max Planck Institute for the Structure and Dynamics of Matter, Hamburg, Germany}
\affiliation{Departamento Fisica de Materiales, Universidad del Pa\'is Vasco, Av. Tolosa 72, E-20018, San Sebastian, Spain}
\affiliation{Center for Computational Quantum Physics, The Flatiron Institute, 162 Fifth Avenue, New York, NY 10010, USA}

\author{Majed Chergui$^\dagger$}
\affiliation{Laboratory of Ultrafast Spectroscopy, ISIC and Lausanne Centre for Ultrafast Science (LACUS), \'Ecole Polytechnique F\'{e}d\'{e}rale de Lausanne, CH-1015 Lausanne, Switzerland}

\date{\today}

\begin{abstract}
Elucidating the carrier density at which strongly bound excitons dissociate into a plasma of uncorrelated electron-hole pairs is a central topic in the many-body physics of semiconductors. However, there is a lack of information on the high-density response of excitons absorbing in the near-to-mid ultraviolet, due to the absence of suitable experimental probes in this elusive spectral range. Here, we present a unique combination of many-body perturbation theory and state-of-the-art ultrafast broadband ultraviolet spectroscopy to unveil the interplay between the ultraviolet-absorbing two-dimensional excitons of anatase TiO$_2$ and a sea of electron-hole pairs. We discover that the critical density for the exciton Mott transition in this material is the highest ever reported in semiconductors. These results deepen our knowledge of the exciton Mott transition and pave the route toward the investigation of the exciton phase diagram in a variety of wide-gap insulators. 
\end{abstract}


\maketitle


One of the major intellectual advancements in modern condensed matter physics has been the formulation of the insulator-to-metal transition, first given by Mott~\cite{mott1968metal}. In his description, Mott proposed that increasing the carrier density in an insulator leads to the screening of the underlying Coulomb potential. Above a critical density known as the Mott density ($n_{M}$), bound states cease to exist and the material eventually turns into a metal. The consequences of this theoretical prediction have been far-reaching, revealing unprecedented insights into the properties of solids such as highly-doped band semiconductors~\cite{klingshirn2012semiconductor}, excitonic insulators \cite{moskalenko2000bose}, and strongly correlated electron systems \cite{khomskii2014transition}.

In the case of a band semiconductor, the simplest bound states are represented by excitons, collective excitations of electron-hole (e-h) pairs coupled via the long-range Coulomb interaction. Increasing the carrier density up to the degenerate limit reinforces the fermionic coupling among the electrons and holes, ultimately resulting in the dissociation of the bound states above $n_{M}$. The Mott criterion predicts that the transition occurs if $k_S$~$a_B$~$\simeq$~1.19, where $k_S$ is the critical screening length at which bound states nominally disappear and $a_B$ is the exciton Bohr radius. More refined theoretical analysis \cite{mahan1967excitons, moskalenko2000bose, guerci2019exciton} and extensive experimental work \cite{suzuki2012exciton, sekiguchi2015excitonic, shahmohammadi2014biexcitonic, palmieri, sekiguchi2017anomalous, omachi2013observation, almand2014quantum} have instead revealed a rich phase diagram of exotic states persisting above $n_{M}$. Notable examples include robust excitonic and biexcitonic correlations \cite{suzuki2012exciton, sekiguchi2015excitonic, shahmohammadi2014biexcitonic}, emergent Mahan excitons \cite{mahan1967excitons, palmieri}, anomalous metallic states \cite{sekiguchi2017anomalous}, e-h droplets \cite{omachi2013observation, almand2014quantum}, and possible Bose condensates of photoexcited e-h Cooper pairs \cite{moskalenko2000bose}. Therefore, the identification of $n_{M}$ in semiconductors acquires a crucial importance for discovering hitherto-unobserved phenonena and clarifying how excitons react to the large carrier densities present in many optoelectronic devices.

One solid that has recently emerged as a promising platform to explore bound exciton physics is the anatase polymorph of TiO$_2$ \cite{ref:baldini_TiO2, baldini2018phonon, baldini2019exciton}, a material extensively used in light-energy conversion applications \cite{ref:fujishima, ref:oregan} and transparent conducting substrates \cite{furubayashi2005transparent}. This system is an indirect gap insulator (Fig. 1(a)): the valence band (VB) top is close to the X point of the Brillouin zone, whereas the conduction band (CB) bottom lies at the $\Gamma$ point. The optical spectrum is dominated by a prominent direct excitation around 3.80 eV (peak I in Fig. 1(b)), which lies on the tail of indirect interband transitions (similar to bulk transition metal dichalcogenides \cite{molina2013effect}). Since the energy of peak I is significantly lower than the direct quasiparticle gap of 3.98 eV, this transition is a rare type of strongly bound exciton with binding energy ($E_B$) larger than 150~meV \cite{ref:baldini_TiO2}. Such a large $E_B$ stems from the contribution of many single-particle states in building up the exciton wavefunction along the $\Gamma$-Z symmetry direction, where the VB and CB have almost parallel dispersion (violet arrows in Fig.~1(a)). Calculations reveal that these excitons have an intermediate character between the Wannier and the Frenkel limit, and are characterized by a two-dimensional (2D) wavefunction in the three-dimensional (3D) lattice (inset to Fig. 1(b)) \cite{ref:baldini_TiO2}. The large $E_B$ makes them particularly immune to perturbations, such as temperature or the scattering at impurities and defects. As a result, these collective excitations manifest themselves also in the room temperature (RT) absorption spectrum of the defect-rich nanoparticles used in typical light-conversion applications \cite{ref:baldini_TiO2, baldini_dye, baldini2018phonon}. However, the extent to which these excitons persist against a high density of free carriers injected in the bands is yet to be addressed. On a fundamental side, shedding light on this problem would show how different many-body effects conspire to destabilize a 2D bound state in a 3D crystal and establish whether the exciton can form stable polaritonic states in \textit{ad-hoc}-designed microcavities. On the technological side, a deeper knowledge of $n_{M}$ would guide the rational design of effective transparent conducting substrates based on TiO$_2$.

\begin{figure}[t]
	\begin{center}
		\includegraphics[width=\columnwidth]{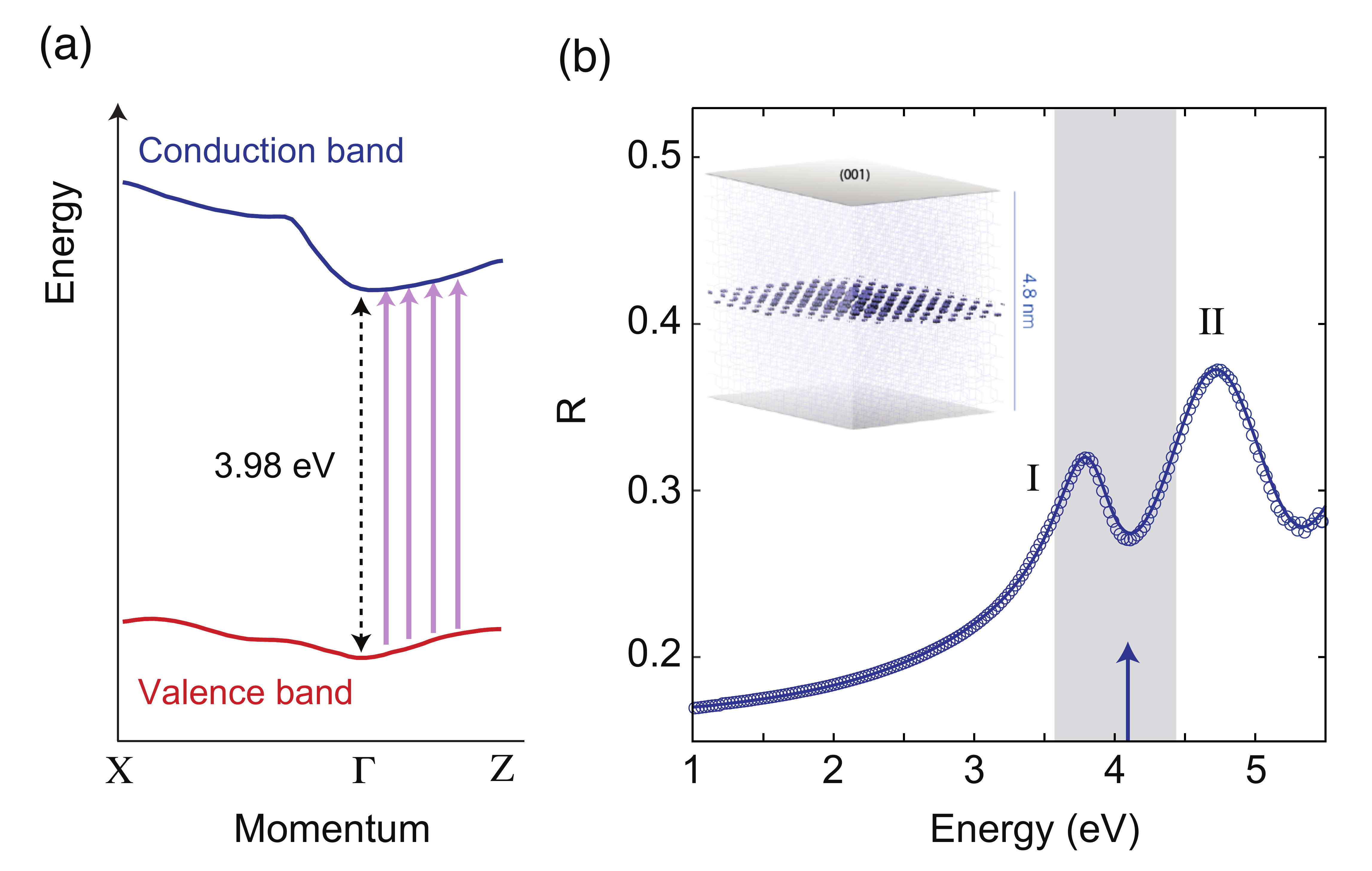}
		\caption{(a) Schematic illustration of the electronic band structure of anatase TiO$_2$, as obtained from $GW$ calculations. The direct gap around the $\Gamma$ point is determined in experiments to be $\sim$ 3.98 eV \cite{ref:baldini_TiO2}. The violet arrows indicate the single-particle states contributing to build up the $a$-axis bound exciton transition. (b) Reflectivity spectrum of anatase TiO$_2$ at RT with the light electric field polarized along the $a$-axis. Peak I is the 2D bound exciton, whereas peak II is a high-energy resonant exciton. The data, measured by spectroscopic ellipsometry, and their assignment are obtained from Ref. \cite{ref:baldini_TiO2}. The pump photon energy of 4.10 eV used for the pump-probe experiment is indicated by the blue arrow and the probed region is highlighted as a grey shaded area. The inset shows the wavefunction of the bound 2D exciton around 3.79 eV. The isosurface representation shows the electronic configuration when the hole of the considered excitonic pair is localized close to one oxygen atom. The coloured region represents the excitonic squared modulus wavefunction.}
		\label{fig:Fig1}
	\end{center}
\end{figure}

Nevertheless, addressing this problem poses considerable challenges to currently-available theoretical and experimental techniques. Theoretically, one should build realistic models of the material's electronic and optical properties that account for the plethora of many-body processes induced by the free carriers. In this respect, many-body perturbation theory has revolutionalized the description of the equilibrium electrodynamical properties of materials \cite{onida2002electronic}, but its application to doped semiconductors is still in its infancy \cite{schleife2011optical, pogna2016photo}. Experimentally, one would need an accurate method to inject a high density of free carriers and monitor the modification of the exciton optical lineshape. This task cannot be accomplished by measuring the optical response of the material upon chemical doping, since the dopant-induced inhomogeneous broadening \cite{makino2002optical, feneberg2014band} and possible electron-electron correlations \cite{yong2016emerging} would mask the effects induced by the free-carrier density. A more powerful approach relies on photodoping the crystal out of equilibrium using an above-gap laser pulse and mapping the optical response around the exciton resonance with sub-picosecond time resolution. Unlike its steady-state analogue, this technique allows for disentangling the contributions of different optical nonlinearities on the exciton peak, based on their characteristic timescale. In TiO$_2$, this would require the simultaneous generation of intense pump and broadband probe pulses covering the elusive near-to-mid UV range (3.20-4.50 eV), a technology that has long been limited by constraints in nonlinear optical conversion schemes \cite{shih2011ultrafast, ref:aubock, west2012two, borrego2019two}.

In this Letter, we set a first milestone toward the determination of $n_{M}$ for a bound exciton absorbing UV light. We achieve this in anatase TiO$_2$ single crystals via a unique combination of many-body perturbation theory and state-of-the-art ultrafast broadband UV spectroscopy. We reveal that the 2D excitons are stable bound quasiparticles in the material at least up to a giant carrier density of $\sim$ 5 $\times$ 10$^{19}$ cm$^{-3}$ at RT. Our results show that the bound states in TiO$_2$ are among the most robust excitons reported so far and open intriguing perspectives for the study of many-body e-h correlations in a wide class of insulators that have remained inaccessible.

As a first step in our study, we explore theoretically the interplay between the bound 2D excitons of anatase TiO$_2$ and free carriers by computing the $GW$ band structure within the frozen lattice approximation in a uniformly electron-doped crystal. Thereafter, we obtain the optical response in the presence of e-h correlations by solving the Bethe-Salpeter equation at the different doping levels \cite{hedin1965new, onida2002electronic}. More details are provided in the Supplementary Material (SM). A thorough comparison between the theoretical and experimental response at zero doping was given in Ref. \cite{ref:baldini_TiO2}. In Fig. 2(a) we only focus on the doping ($n$) dependence, which shows a strongly nonlinear response of the single-particle gap (dashed vertical lines) and the optical spectra with $n$. In particular, we find that the quasiparticle gap and exciton absorption of the system do not change between $n$ =~0~cm$^{-3}$ and $n$~$\sim$~1.4~$\times$~10$^{19}$~cm$^{-3}$. At $n$~=~14~$\times$~10$^{19}$~cm$^{-3}$, the exciton peak blueshifts by $\sim$ 50 meV, a value that is larger than the carrier-induced blueshift of the quasiparticle gap ($\sim$~20 meV). As a result, $E_B$ is weakened by $\sim$ 30 meV. Increasing $n$ to 35 $\times$ 10$^{19}$ cm$^{-3}$ results in an abrupt and large redshift of the quasiparticle gap due to band-gap renormalization (BGR). Here, the quasiparticle gap overlaps the exciton peak energy, signaling the occurrence of the Mott transition. However, even if bound states cease to exist above $n_M$, excitonic correlations still persist in the form of a resonant exciton that shapes the optical response \cite{baldini2017anomalous}. Further increasing $n$ results in a substantial smearing of this resonant exciton and in the shrinking of the quasiparticle gap. The complete dependence of $E_B$ on $n$ is shown in Fig. 2(b). From this plot, we estimate that the Mott transition occurs at a surprisingly high value of $n_M$ $\sim$ 35 $\times$ 10$^{19}$ cm$^{-3}$. 

\begin{figure}[t]
	\begin{center}
		\includegraphics[width=\columnwidth]{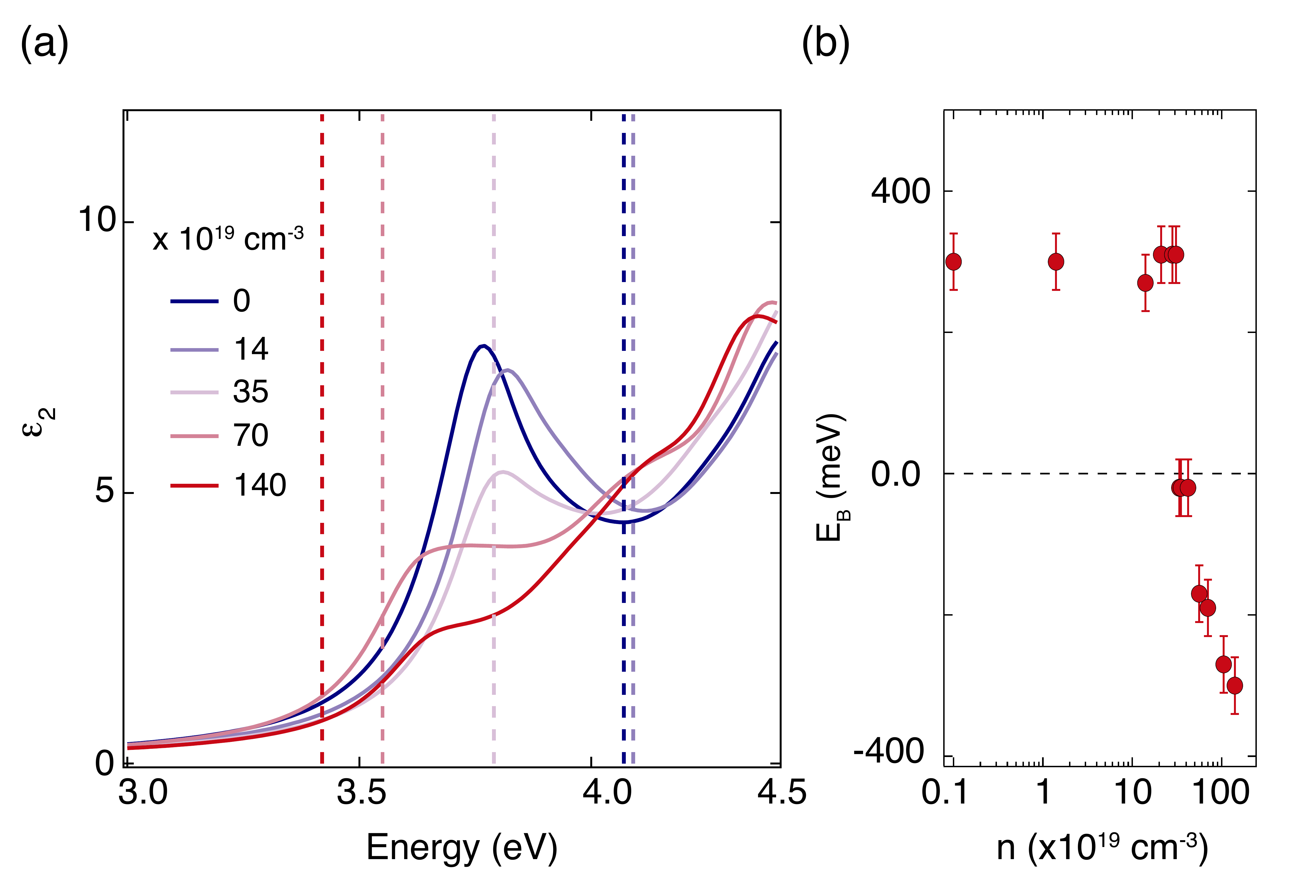}
		\caption{(a) Imaginary part of the $a$-axis dielectric function of anatase TiO$_2$, calculated by solving the Bethe-Salpeter equation for pristine and $n$-doped TiO$_2$. The optical response of the $n$-doped crystal with $n$ = 1.4 $\times$ 10$^{19}$ cm$^{-3}$ (not shown) overlaps almost completely that of the pristine case, indicating that this doping level does not affect the exciton peak energy. Only when $n$ $>$ 10 $\times$ 10$^{19}$ cm$^{-3}$, does the exciton blueshift. The vertical lines represent the quasiparticle gap energy at each doping level. The Mott transition occurs at $n_{M}$ $\sim$ 35 $\times$ 10$^{19}$ cm$^{-3}$, i.e. when the quasiparticle gap matches the exciton peak energy. (b) Doping dependence of $E_B$, estimated from the energy difference between the quasiparticle gap energy and the exciton peak energy. An abrupt change is observed around $n$ $>$ 10 $\times$ 10$^{19}$ cm$^{-3}$ and the bound states are lost at $n_M$.}
		\label{fig:Fig4}
	\end{center}
\end{figure}

Next, we investigate this finding experimentally using ultrafast spectroscopy with a near-to-mid-UV continuum probe. The description of the experimental methods is given in the SM. Our goal is to map the response of the bound exciton in TiO$_2$ upon illumination with an intense laser pulse centered around 4.10 eV (blue arrow in Fig. 1(b)). This photon energy lies above the exciton peak and thus excites uncorrelated e-h pairs in the solid. We set the incident fluence to the maximum value that our state-of-the-art apparatus can deliver and we carefully convert it into a density of photoexcited carriers (see the SM for the estimate of the uncertainties). We obtain $n$~$\sim$~5~$\times$~10$^{19}$~cm$^{-3}$, i.e. below the theoretically-predicted $n_M$ but sufficiently high compared to the density at which excitons dissociate in most solids \cite{amo2006interplay, palmieri, klingshirn2007room, watanabe2000nonlinear, choi2001ultrafast}. Subsequently, we monitor the relative changes in the material reflectivity ($\Delta$R/R) over a broad spectral range covering the bound exciton feature (grey shaded area in Fig. 1(b)). Depending on the spectral extension of our probe pulse, the time resolution of the set-up varies between 80 fs and 1 ps \cite{ref:aubock}.

Figure 3(a) displays the color-coded map of $\Delta R/R$ as a function of the probe photon energy and pump-probe time delay. To allow for a broadband detection between 3.60 and 4.40 eV, the time resolution of the set-up is set at 700 fs. We observe that the signal is positive above $\sim$ 3.95~eV and negative below this energy. The zero-crossing point varies with time, suggesting a change in the peak position and linewidth of the exciton peak. To visualize these changes, we reconstruct the pump-induced temporal evolution of the material's reflectivity by combining our steady-state and time-resolved optical data. The results, shown in Fig. 3(b), indicate that upon photoexcitation the exciton band decreases its absolute reflectivity, its linewidth broadens, and it shifts to the blue. The wide spectral region covered by this measurement enables us to perform a quantitative analysis of the reflectivity data and obtain the corresponding absorption spectra at different time delays. To this aim, we fit the steady-state optical data with a Lorentz model, as shown by the solid line in Fig. 1(b). Thereafter, we describe the pump-induced changes of the reflectivity spectrum through the variation of the bound exciton parameters (details are given in the SM). Iterating the fit at each time delay yields the time-dependent absorption coefficient $\alpha$($\omega$, t) (Fig. 3(c)), as well as the time evolution of the exciton oscillator strength, linewidth, and peak energy (Fig. 3(d)-(f)). At the present photoexcited carrier density, we find that the exciton oscillator strength decreases by only $\sim$ 5$\%$ (Fig. 3(d)) within our time resolution and recovers with a bi-exponential trend with timescales of 0.85 $\pm$ 0.37 ps and 33 $\pm$ 7 ps. In contrast, a different temporal behavior is shared by the exciton peak energy (Fig. 3(e)), and linewidth (Fig. 3(f)). In particular, the exciton peak energy increases by $\sim$ 35 meV and recovers on timescales of 20 $\pm$ 4 ps and 250 $\pm$ 111 ps. This suggests that the same optical nonlinearity causes both the linewidth and peak energy increase.

\begin{figure*}[t]
	\begin{center}
		\includegraphics[width=1.7\columnwidth]{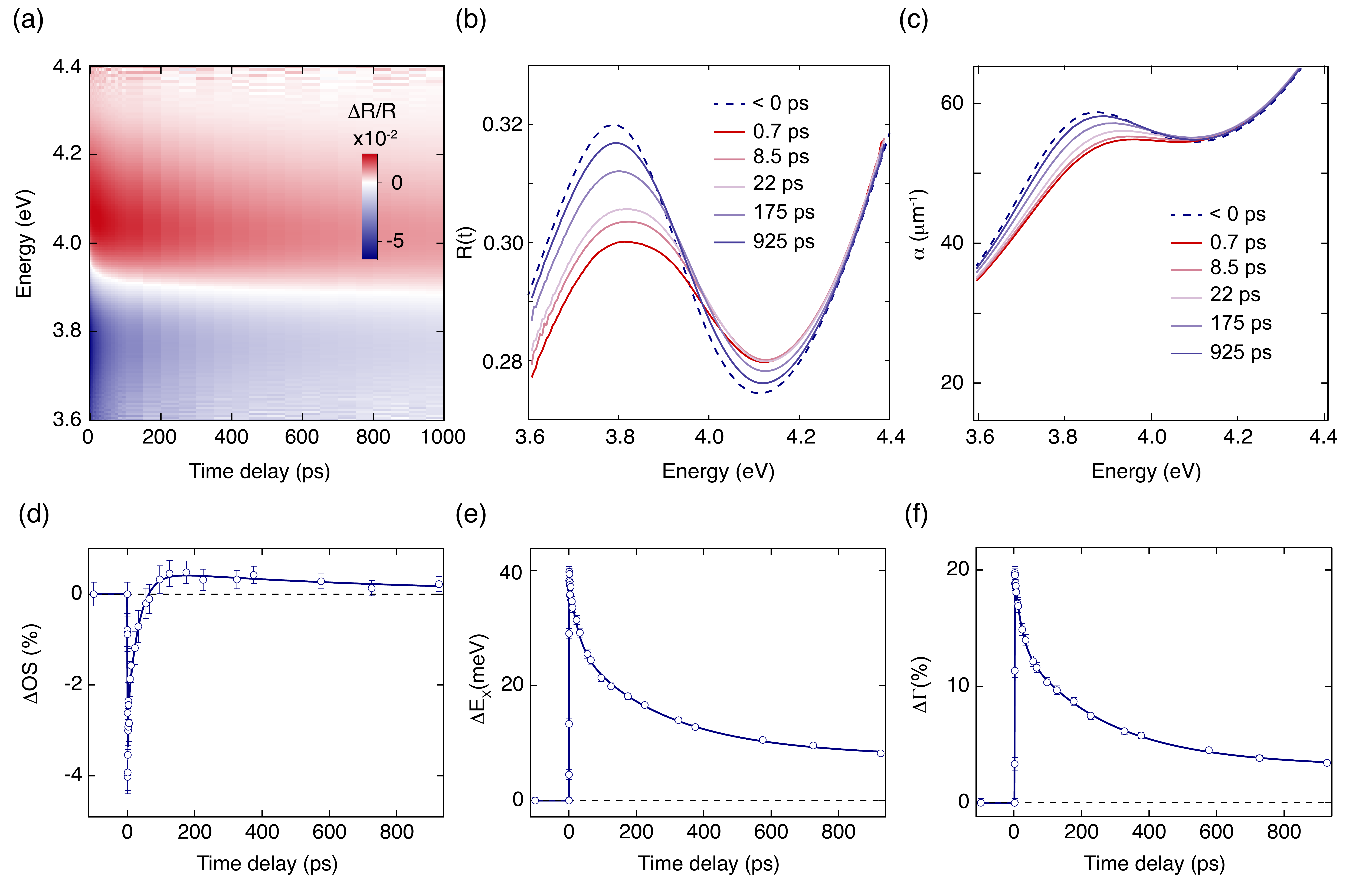}
		\caption{Spectral evolution of R($\omega$, t), obtained by combining the steady-state R($\omega$) and the $\Delta$R/R($\omega$, t) data in the time window 700 fs - 1 ns. The R of the unperturbed system is displayed as a dashed blue line. (b) Spectral evolution of $\alpha$($\omega$,~t), obtained as a result of the Lorentz fit in the time window 700 fs - 1 ns. The $\alpha$ of the unperturbed system is displayed as a dashed blue line. (c-e) Temporal evolution of (c) the oscillator strength, (d) the linewidth, and (e) the peak energy of the bound exciton upon photoexcitation.}
		\label{fig:Fig3}
	\end{center}
\end{figure*}

The spectro-temporal analysis of the exciton lineshape allows us to disentangle the single-particle and many-body effects participating in the exciton optical nonlinearities \cite{ref:haug}. In particular, the presence of electron/hole populations inside the bands partially block the transitions contributing to the exciton state (indicated by the violet arrows in Fig. 1(a)). As a result, this phase-space filling (PSF) of the relevant single-particle states causes a decrease in the exciton oscillator strength. Owing to the characteristic electronic structure of anatase TiO$_2$ (Fig.~1(a)), the photoexcited electrons relax to the bottom of the CB at $\Gamma$, whereas the holes to the top of the VB close to X. As such, the PSF contribution to the exciton spectral changes arises exclusively due an electron population close to the bottom of the CB. Moreover, the increased broadening of the exciton linewidth is a manifestation of long-range Coulomb screening (CS), as the photoexcited carrier density reduces the exciton lifetime \cite{baldini_dye}. Finally, the origin of the exciton blueshift over time deserves deeper attention. Three known optical nonlinearities can cause a shift in an exciton peak upon above-gap illumination, namely BGR, CS, and PSF \cite{ref:haug, ref:schmitt-rink}. BGR leads to a density-dependent shrinkage of the single-particle states and hence a redshift of the exciton feature due to the change in electron-electron and electron-ion interaction. Long-range CS modifies the e-h interaction, resulting in a reduced $E_\mathrm{B}$ and in a blueshift of the exciton peak. Finally, PSF may also contribute to the blueshift of the exciton peak: indeed, the carriers populating the band edges can lead to the apparent shift of the overall absorption edge toward high energies (Burstein-Moss effect) \cite{burstein1954anomalous, moss1954interpretation}. These processes act simultaneously on the exciton peak energy, their relative weights being governed by the material parameters and dimensionality \cite{ref:haug, ref:schmitt-rink}. However, under the present excitation conditions, the exciton peak energy temporal response closely resembles that of its linewidth. This strongly indicates that long-range CS is the dominant nonlinearity behind the exciton blueshift, ruling out PSF and BGR. Therefore, the detected exciton blueshift can be directly correlated with the absolute change in $E_\mathrm{B}$ produced by the photoexcited e-h plasma. Since $E_\mathrm{B}$ changes only by $\sim$ 35~meV at 700~fs, excitons are bound entities at this time delay.

As in TiO$_2$ the intraband carrier relaxation is complete within 50 fs (due to the strong electron-phonon coupling) \cite{bothschafter2013ultrafast, baldini2018clocking}, one may argue that the results obtained at 700~fs are not representative of the stated excitation density. Indeed, recombination mechanisms such as carrier trapping or three-body Auger processes may have already decreased the actual carrier density contributing to the exciton screening in this indirect-gap material. This requires one to resolve a well-defined exciton feature at a time delay close to 50 fs, demonstrating the persistence of the e-h correlations at such a short timescale. Due to the trade-off between time resolution and probe spectral coverage in our set-up, we also demonstrate the stability of the excitonic correlations at 120 fs by resolving their signature over a narrower range in the reflectivity spectrum. The results, reported in Fig. S5, indicate that the exciton peak is not entirely suppressed by the presence of the e-h plasma and that the excitonic correlations are still intact in this highly non-equilibrium phase. 

We believe that the photoexcited carrier density in our experiment lies below the actual value of $n_M$. Persistence of excitonic correlations above $n_M$ have been recently demonstrated in several semiconductors \cite{suzuki2012exciton, sekiguchi2015excitonic, palmieri} and the emergence of Mahan excitons has been invoked \cite{mahan1967excitons, palmieri}. However, in such a scenario, the Wannier exciton feature would be accompanied by the enhancement of the continuum absorption, which is instead absent in our data. Therefore, our result support a scenario in which the actual $n_M$ is larger than 5 $\times$ 10$^{19}$ cm$^{-3}$. Our many-body perturbation theory does not account for finite temperature effects and the presence of quasi-Fermi energies for a nonequilibrium distribution of e-h pairs. Future extensions of our theory to include these effects will refine our theoretically-predicted $n_M$, most likely around lower values. Despite these corrections, we can confidently conclude that the Mott transition occurs in anatase TiO$_2$ at a remarkably-high $n_M$. For comparison, other bulk insulators supporting bound excitons built upon the single-particle states have $n_M$ varying between 7 $\times$ 10$^{16}$ cm$^{-3}$ and 6.4 $\times$ 10$^{18}$ cm$^{-3}$ (see Table S1). In the case of TiO$_2$, such a high $n_M$ can be explained by the robustness of the single-particle gap to the injected carrier density, consistent with our calculations and observations by angle-resolved photoemission spectroscopy \cite{ref:baldini_TiO2}.

In conclusion, our work demonstrates the robustness of the bound excitons in TiO$_2$ and shows the power of ultrafast broadband UV spectroscopy to investigate many-body phenomena involving high-energy excitons and large carrier densities. We envision the application of this method to study a variety of high-energy excitons that strongly couple to the lattice or the spin degrees of freedom, i.e. in perovskite titanates \cite{gogoi2015anomalous, begum2019role} or in antiferromagnetic Mott insulators \cite{rodl2012optical}.

\begin{acknowledgments}
We thank Alexander Steinhoff for insighful discussions, and Simon Moser and Marco Grioni for providing the sample used for this study. We acknowledge support by the Swiss National Science Foundation via the NCCR:MUST, contract 154056 (PNR 70, ``Energy turnaround") and the R'EQUIP contract 206021-157773. We acknowledge financial support from the European Research Council (ERC-2015-AdG-694097) and the Cluster of Excellence Advanced Imaging of Matter (AIM) EXC 2056 - 390715994. The Flatiron Institute is a division of the Simons Foundation. Support by the Max Planck Institute - New York City Center for Non-Equilibrium Quantum Phenomena is acknowledged.\\
\end{acknowledgments}

\noindent * ebaldini@mit.edu\\
$\dagger$ majed.chergui@epfl.ch
\newpage
\clearpage

\setcounter{section}{0}
\setcounter{figure}{0}
\renewcommand{\thesection}{S\arabic{section}}  
\renewcommand{\thetable}{S\arabic{table}}  
\renewcommand{\thefigure}{S\arabic{figure}} 
\renewcommand\Im{\operatorname{\mathfrak{Im}}}

\section{I. First-principles calculations}

\subsection{A. Electronic structure}

The electronic structure of pristine and electron-doped anatase TiO$_2$ was calculated using many-body perturbation theory at the one-shot $GW$ level \cite{hedin1965new}. This approach has been shown to describe accurately the electronic properties of many band semiconductors \cite{onida2002electronic}. In the case of anatase TiO$_2$, the $GW$ electronic band structure and gap size are in excellent agreement with those obtained from angle-resolved photoemission spectroscopy measurements \cite{ref:baldini_TiO2}. 

The system was modelled by using the primitive unit cell of anatase TiO$_2$ with lattice parameters $a$~=~$b$~=~3.79~$\AA$ and $c$ = 9.67 $\AA$ (unit cell volume of 69.58 $\AA^3$). These values were calculated using the generalized gradient approximation (GGA), and are in line with the experimental values. The Brillouin zone was sampled with a 4$\times$4$\times$4 $k$-point grid. We used a total of 2474 conduction bands and a 46 Ry energy cutoff for the computation of the inverse dielectric matrix. An energy cutoff of 46 Ry and 160 Ry was employed for the evaluation of the screened and the bare Coulomb interaction components of the self-energy operator, respectively. All these parameters were systematically and independently increased until the obtained electronic structure was converged within few tens of meV (see Ref. \cite{ref:baldini_TiO2} for a detailed discussion).        

\begin{figure*}[t]
	\begin{center}
		\includegraphics[width=2\columnwidth]{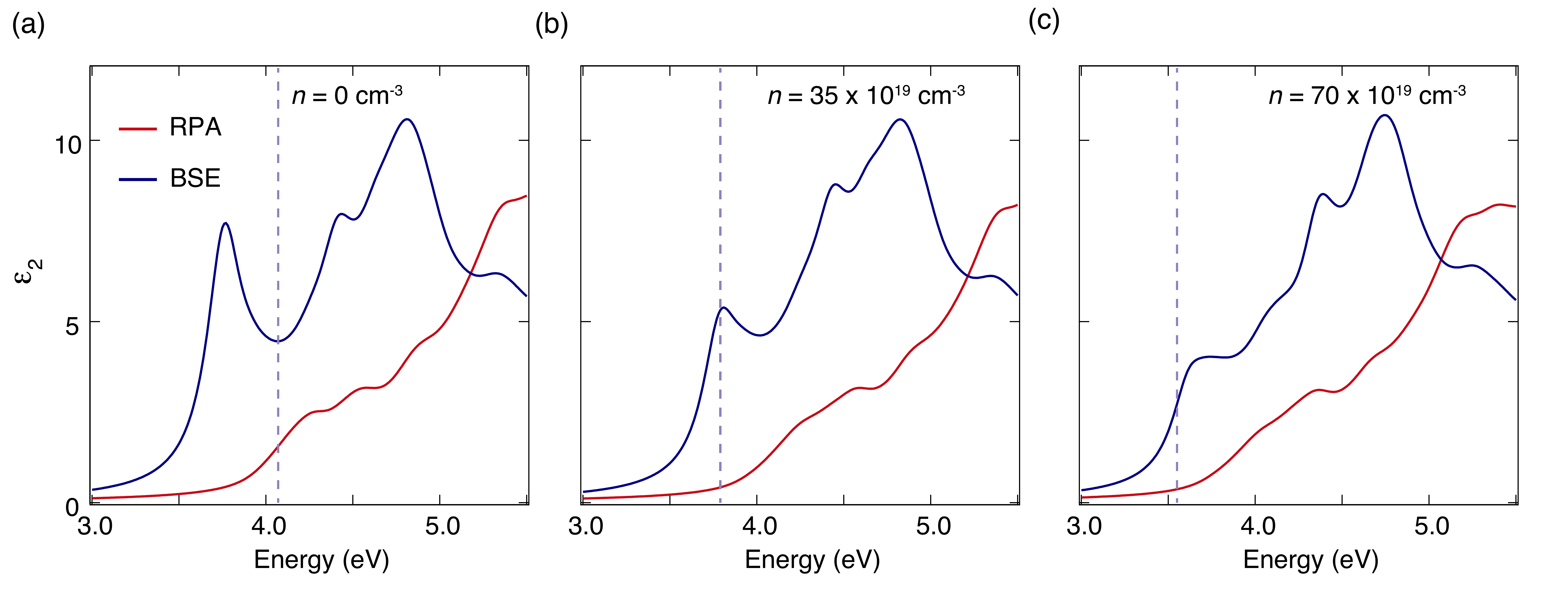}
		\caption{Comparison between the imaginary part of the dielectric function of anatase TiO$_2$ calculated at the RPA-$GW$ level (red lines) and that calculated at the BSE-$GW$ level (blue lines). Three different dopings are shown: (a) $n$ = 0 cm$^{-3}$; (b)~$n$~=~35~$\times$ 10$^{19}$ cm$^{-3}$; (c) $n$ = 70 $\times$ 10$^{19}$ cm$^{-3}$. Only the inclusion of excitonic effects allows us to reproduce the experimental response. In addition, the RPA-$GW$ spectrum signals the onset of the electron-hole continuum. Such an onset is also indicated by the direct quasiparticle gap computed at the $GW$ level, which is represented as a dashed violet vertical line. The excitonic Mott transition occurs when the quasiparticle gap overlaps to the exciton peak energy and the excitons cease to be bound.}
		\label{fig:FigS1}
	\end{center}
\end{figure*}

\subsection{B. Optical response}

We calculated the optical spectra with and without electron-hole correlations, relying on the computed $GW$ quasiparticle energies. In the case with no electron-hole correlations, we used the random phase approximation (RPA) and obtained a featureless optical spectrum that quantifies the contribution of the electron-hole continuum. In contrast, solving the Bethe-Salpeter equation (BSE) allowed us to capture the excitonic features observed in spectroscopic ellipsometry measurements both qualitatively and quantitatively  \cite{ref:baldini_TiO2}. To solve the BSE, we employed a 16$\times$16$\times$16 $k$-point grid and included the 6 topmost valence bands and 6 lowest conduction bands. In the plot of the optical spectra, we applied a Lorentzian broadening with an energy width of 120 meV. All $GW$ and BSE calculations were performed with the BerkeleyGW package \cite{deslippe2012berkeleygw}. Representative RPA and BSE results at different doping levels are shown in Fig. S1. In all three cases, the most precise method to evaluate the rise of the electron-hole continuum for the direct excitations relies on the estimate of the $GW$ direct quasiparticle gap (indicated by the dashed violet vertical line).

\subsection{C. Doping dependence}

To simulate the effects produced by a finite density of free carriers in the system, we performed BSE-$GW$ calculations for electron doping values between 10$^{-3}$ and 10$^{-1}$ electrons per unit cell. This corresponds to electron dopings between 1.4 $\times$ 10$^{19}$ cm$^{-3}$ and 1.4 $\times$ 10$^{21}$ cm$^{-3}$. The characteristic band structure of anatase TiO$_2$ ensures that electron doping of the conduction band can mimic the effects that photodoping of uncorrelated electron-hole (e-h) pairs produces on the exciton response of the material. This can be observed in Fig. 1(a) of the main text. The top of the valence band in anatase TiO$_2$ resides close to the X point of the Brillouin zone, whereas the bottom of the conduction band is at the $\Gamma$ point. As such, the material is an indirect bandgap semiconductor. The single-particle states contributing to the exciton wavefunction are located along the $\Gamma$-Z symmetry line in the Brillouin zone, as indicated by the violet arrows. Photoexcitation of e-h pairs in our pump-probe experiments leads to the rapid cooling of the electrons to the the conduction band minimum at $\Gamma$ and of the holes to the top of the valence band. The effect of the free-carrier density on the exciton comes solely from the electron population at $\Gamma$. Thus, our calculations for $n$-doped TiO$_2$ provide a good account of the effects produced by the uncorrelated e-h pairs at time delays longer than the cooling time ($>$~50~fs).

\subsection{D. Determination of the Mott transition}

Here we describe how we estimate theoretically the exciton Mott density ($n_M$) of anatase TiO$_2$. For a semiconductor/insulator such as anatase TiO$_2$, $n_M$ is defined as the carrier density at which the exciton binding energy ($E_B$) equals zero. Under these circumstances, the exciton is no longer bound and excitonic correlations can persist only in the form of resonant enhancements of the absorption spectrum.

We follow a rigorous approach to determine $E_B$. First, we calculate the single-particle band structure of the material at the $GW$ level of theory for different values of the carrier density. Afterwards, we compute the exciton energy, $E_{exc}$, for each case by solving the BSE on top of the $GW$ results, and identify the single-particle states contributing to the exciton. This allows us to estimate the value of the quasiparticle gap ($E_{qp}$) in the region of the Brillouin zone that builds up the exciton wavefunction. Finally, $E_B$ is given by $E_{qp} - E_{exc}$. In Fig. 2(a), for each excess carrier density, $E_{qp}$ is indicated by a vertical dashed line the same color as the optical spectrum for that density. $E_{exc}$ is simply the energy at the exciton peak in the spectrum. When $n$ = 0 cm$^{-3}$ (pristine anatase case in dark blue), $E_B$ takes its largest value. As we increase $n$, $E_B$ starts decreasing in value, and goes to zero for $n_M$ = 35 $\times$ 10$^{19}$ cm$^{-3}$ (see Fig. 2(b) in main text). Excitonic correlations persist even above $n_M$ in the form of a resonant exciton \cite{baldini2017anomalous}.

\section{II. Single crystal growth and characterization}

High-quality single crystals of anatase TiO$_2$ were produced by a chemical transport method from anatase powder and NH$_4$Cl as transport agent, similar to the procedure described in Ref. \cite{berger1993growth}. In detail, 0.5 g of high-purity anatase powder were sealed in a 3 mm thick, 2 cm large and 20 cm long quartz ampoule together with 150~mg of NH$_4$Cl, previously dried at $60\,^{\circ}\mathrm{C}$ under dynamic vacuum for one night, and 400 mbar of electronic grade HCl. The ampoules were placed in a horizontal tubular two-zone furnace and heated very slowly to $740\,^{\circ}\mathrm{C}$ at the source, and $610\,^{\circ}\mathrm{C}$ at the deposition zone. After two weeks, millimeter-sized crystals with a bi-pyramidal shape were collected and cut into rectangular bars (typically 0.8 $\times$ 0.6 $\times$ 0.15 $\mathrm{mm^3}$).

\section{III. Experimental set-up}

The ultrafast optical experiments were performed using a novel set-up of tunable UV pump and broadband UV probe, described in detail in Ref. \cite{ref:aubock}. A 20 kHz Ti:Sapphire regenerative amplifier (KMLabs, Halcyon + Wyvern500), providing pulses at 1.55 eV, with typically 0.6 mJ energy and around 50 fs duration, pumped a noncollinear optical parametric amplifier (NOPA) (TOPAS white - Light Conversion) to generate sub-90 fs visible pulses (1.77 - 2.30 eV range). The typical output energy per pulse was 13 $\mu$J. Around 60\% of the output of the NOPA was used to generate the narrowband pump pulses. The visible beam, after passing through a chopper, operating at 10 kHz and phase-locked to the laser system, was focused onto a 2-mm thick $\beta$-barium borate (BBO) crystal for nonlinear frequency doubling. The pump photon energy was controlled by the rotation of the crystal around the ordinary axis and could be tuned in a spectral range up to $\sim$0.9 eV ($\sim$60 nm) wide. For our purpose, the pump photon energy was set at 4.10 eV, in order to selectively excite uncorrelated e-h pairs above the first excitonic peak of anatase TiO$_2$. The remaining NOPA output was used to generate the broadband UV probe pulses with $\sim$1.3 eV ($\sim$100 nm) bandwidth through an achromatic doubling scheme.

To study the anatase TiO$_2$ single crystals, the set-up was used in the reflection geometry. The specimens were mounted on a rotating sample holder, in order to explore the transient reflectivity ($\Delta R/R$) along the desired crystalline axis. Pump and probe pulses, which have the same polarization, were focused onto the sample, where they were spatially and temporally overlapped. The spot size of the pump and the probe were 150 $\mu$m and 80 $\mu$m full-width at half-maximum (FWHM) respectively, resulting in a homogeneous illumination of the probed region. The portion of the probe beam reflected by the surface of the crystal was detected and the time evolution of the difference in the UV probe reflection with and without the pump pulse reconstructed. After the sample, the reflected probe was focused in a multi-mode optical fiber (100 $\mu$m), coupled to the entrance slit of a 0.25 m imaging spectrograph (Chromex 250is). The beam was dispersed by a 150 gr/mm holographic grating and imaged onto a multichannel detector consisting of a 512 pixel complementary metal-oxide-semiconductor (CMOS) linear sensor (Hamamatsu S11105, 12.5 $\times$ 250~$\mu$m pixel size) with up to 50 MHz pixel readout, so the maximum read-out rate per spectrum (almost 100 kHz) allowed us to perform shot-to-shot detection easily. The time resolution varied between 1 ps and 80 fs depending on the spectral coverage of the probe pulse. All the experiments were performed at room temperature. 

\begin{figure*}[t]
	\begin{center}
		\includegraphics[width=1.4\columnwidth]{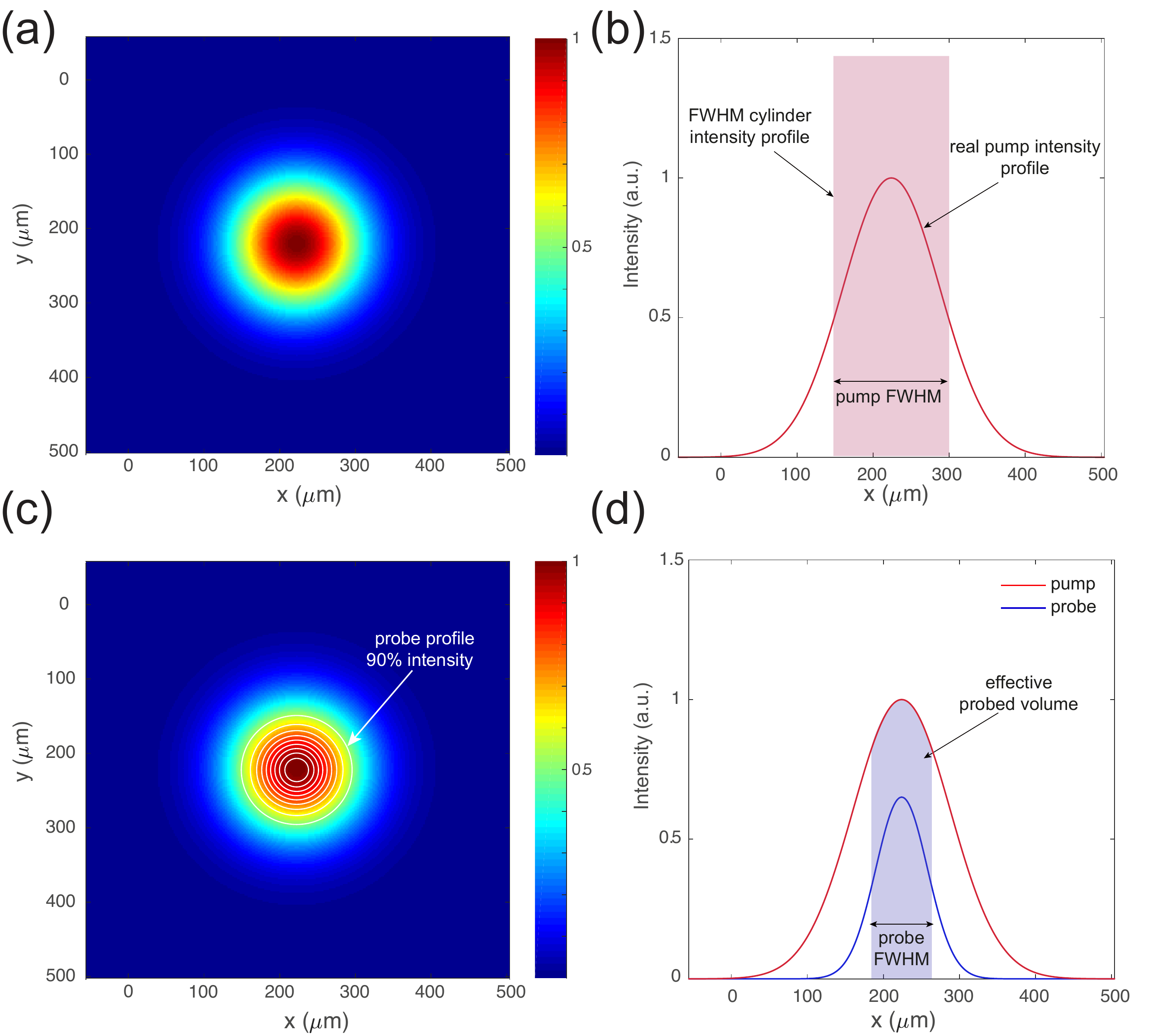}
		\caption{(a) Simulation of the gaussian excitation beam used in our experiment (FWHM = 150 $\mu$m). (b) Profiles of the two-dimensional intensity distributions of the gaussian beam of panel (a) (red curve) and the one obtained by approximation of the gaussian beam with a cylinder. (c) Simulated gaussian intensity distributions of the pump (color-coded) and probe (white contour lines) beams used in the experiment. (d) Profiles of the two-dimensional intensity distributions of the pump and probe beams of panel (c), showing that in these experimental conditions the probed area coincides with the most intense part of the gaussian.}
		\label{fig:FigS2}
	\end{center}
\end{figure*}

\section{IV. Estimate of the photoexcited carrier density}

\label{SN:eh_density}
To explore the ultrafast optical response of TiO$_2$ single crystals in the high-density regime, it is crucial to accurately estimate the experimental e-h density, $n_{e-h}$, created by the pump pulse. This quantity can be expressed as
\begin{equation}\label{eq:n}
	n_{e-h}=(1-R)\frac{F}{h\nu\lambda_\text{p}},
\end{equation}
where $F$ is the pump fluence, $h\nu$ is the pump photon energy, $\lambda_\text{p}=1/\alpha$ the light penetration depth in the material, and $R$ is the reflectivity of the sample. All parameters are evaluated at the pump photon energy (4.10 eV). The uncertainty on $n_{e-h}$ can be estimated by propagating the uncertainty in the variables entering Supplementary Equation~(\ref{eq:n}), namely the absorption/reflection coefficients of the sample and the laser parameters. Moreover, as the choice of the excitation volume geometry is arbitrary, we discuss the approximations introduced for the calculation of the excitation spot size.\\

\begin{figure*}[t]
	\begin{center}
		\includegraphics[width=1.7\columnwidth]{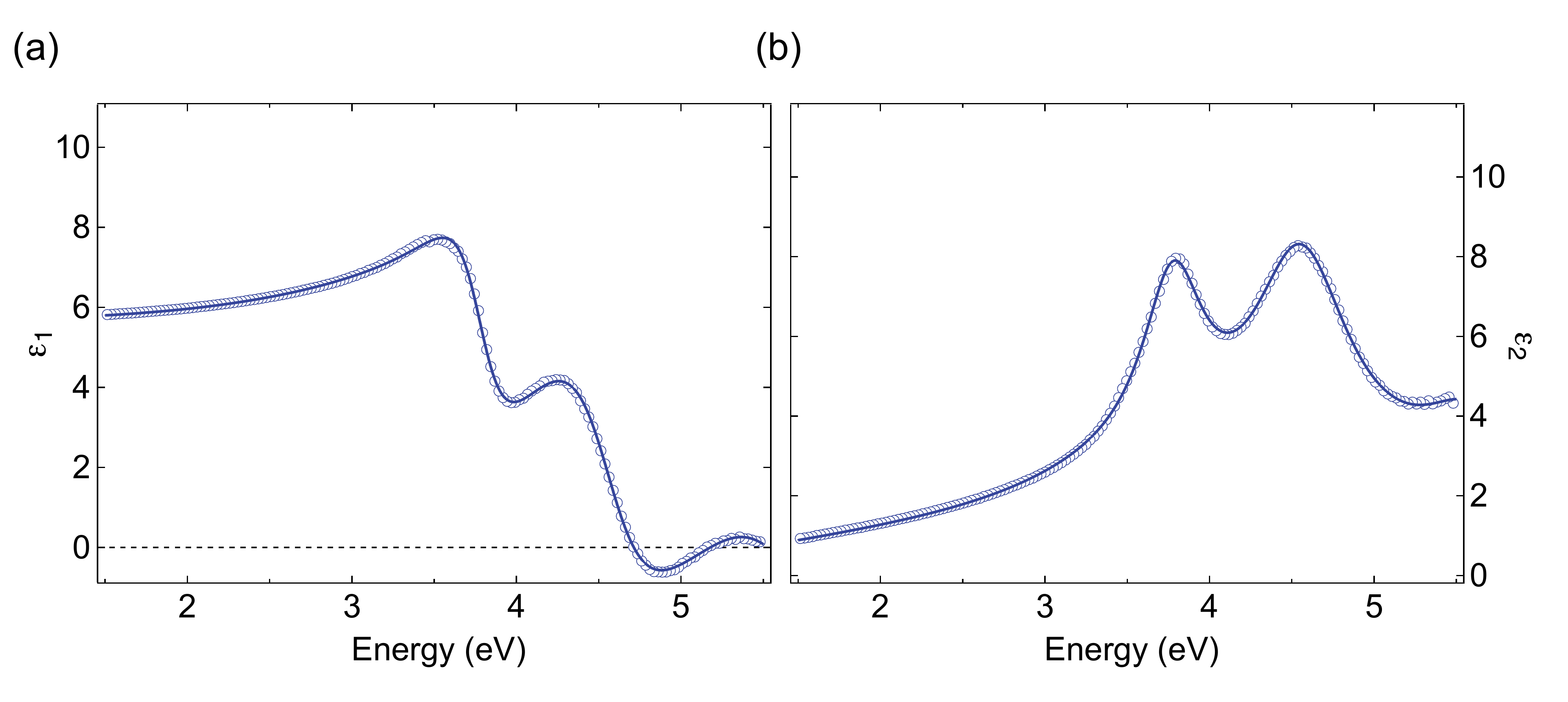}
		\caption{(a) Real ($\epsilon_1(\omega)$) and (b) imaginary ($\epsilon_2(\omega)$) parts of the complex dielectric function of anatase TiO$_2$, measured at room temperature with spectroscopic ellipsometry (blue dots) along the crystallographic a-axis. Solid blue lines show the simultaneous fits of both measured data set to the same Lorentz model.}
		\label{fig:FigS3}
	\end{center}
\end{figure*}

\textbf{Absorption and reflection coefficients:} The estimates of the absorption coefficient, as well as the amount of reflection from the sample surface, are based on measured data of spectroscopic ellipsometry, which is the most accurate experimental technique currently available to determine the real and imaginary parts of the dielectric function for any insulator above its fundamental gap. Our spectroscopic ellipsometry data are shown in Ref. \cite{ref:baldini_TiO2} and the absorption/reflectance spectra obtained directly from the measured optical quantities without the need of a Kramers-Kronig analysis. Since the error on the measured ellipsometry angles $\mathit{\Psi}$ and $\mathit{\Delta}$ is less than 0.1$\%$, the uncertainty in $R$, $\alpha$, and $\lambda_\text{p}$ remains well below 1$\%$.\\

\textbf{Incident laser fluence:} The average incident laser fluence $F$ (measured in $\mu$J/cm$^{2}$) is defined as $F = P/(r \cdot A)$, where $P$ is the laser power, $r$ the repetition rate of the laser system, and $A$ the laser spot size. The impinging laser power $P$ is measured accurately by using an ultraviolet-extended ultra-sensitive photodiode. For our photodiode, the uncertainty of the power measurement at a photon energy of 4.10 eV is equal to $\delta P = \pm 4\%$. The measurement of the spot size is performed using a camera-based beam profiling system consisting of a camera and analysis software. Here, the uncertainty in the width measurement is $\delta w = \pm 2\%$. Error propagation from $w$ and $P$ yields to an uncertainty in the incident laser fluence of $\delta F = \sqrt{2\delta w + \delta P} \approx 5\%$.\\


\textbf{Choice of the excitation volume:}
Although the uncertainty in $n_{e-h}$ related to the laser parameters and absorption/reflection of the sample remains below $6\%$, an additional source of variation in the estimated e-h density originates from the arbitrary choice of the excitation volume. In agreement with the approach used in the literature, as excitation volume we consider a cylinder whose area $A$ corresponds to the the laser spot size on the sample and whose height is equal to the light penetration depth $\lambda_\text{p}$ (\emph{i.e.}, the depth at which the intensity of the radiation is decreased by approximately $1/e=37\%$ of its initial value). This choice of $\lambda_\text{p}$ is justified by the similarity between the absorption coefficient of the pump and probe in the explored spectral range; thus, we can reasonably consider the same penetration depth for the pump and probe energies. On the other side, the choice of $A$ deserves more attention. In Fig. S2(a), we simulate the gaussian excitation beam used in our experiment. Conventionally, the spot size diameter is approximated by the FWHM of the gaussian profile
\begin{equation}
	A = \pi \left(\frac{\text{FWHM}}{2}\right)^2,
\end{equation}
Figure S2(b) shows the results of such an approximation by comparing the profiles of the two-dimensional intensity distributions of the gaussian beam (red curve) and the FWHM cylinder. In the former case, since the total intensity (\textit{i.e.} the total volume under the gaussian surface) is contained in a smaller base, the avarage peak intensity is almost 1.5 times the peak intensity of the gaussian. However, the validity of this approximation depends on the relative size of the probe beam with respect to the pump, since a small probe will be able to probe \emph{locally} the photoexcited surface and provide a more precise estimation of the e-h density. This concept is illustrated in Figs. S2(c,d), where we compare the simulated profiles of the pump and probe beams used in the experiment. When the two laser beams are perfectly overlapped, being the probe much narrower than the pump, the probed area will coincide with the most intense part of the gaussian, whose intensity is almost 1.5 times smaller than the one resulting from the FWHM approximation. With the latter being the approximation we use in our study, the declared excitation densities may exceed the actual densities by a factor 1.5-2.

Therefore, by considering all sources of uncertainty presented above, we conclude that the densities reported in the manuscript will differ from the actual densities by a factor no greater than 1.5-2. Such an uncertainty does not influence the discussion and conclusion of our study.

\section{V. Lineshape analysis}

In this Section, we describe the lineshape analysis performed to track the relevant exciton parameters in our pump-probe experiment. As a first step, we modeled the steady-state complex dielectric function with a set of Lorentz oscillators. The real ($\epsilon_1(\omega)$) and imaginary ($\epsilon_2(\omega)$) parts of the dielectric function were measured directly with spectroscopic ellipsometry \cite{ref:baldini_TiO2}. Figure~S3(a,b) shows the experimental traces (blue dotted lines) and the results of the Lorentz model (blue solid lines), indicating the high accuracy of our fit. The fit function comprises four Lorentz oscillators (accounting for the indirect gap transition at 3.53 eV, the bound exciton at 3.77 eV, the resonant exciton at 4.55 eV, and an interband transition that captures the high-energy response), as well as a background tail due to defect-assisted transitions and residual scattered light from surface inhomogeneities. Figure S4 displays the decomposition of the $\epsilon_2(\omega)$ spectrum into the different Lorentzian contributions. Next, we used the modeled $\epsilon_1(\omega)$ and $\epsilon_2(\omega)$ to determine the reflectivity $R(\omega)$ using standard electrodynamical formulas \cite{dressel2002electrodynamics}. The final result was compared to the experimental reflectivity spectrum $R(\omega)$. The latter was also determined by combining the experimental $\epsilon_1(\omega)$ and $\epsilon_2(\omega)$ measured by ellipsometry. The accuracy of the fit can be observed in Fig. 1(b) in the main text.

\begin{figure}[b]
	\begin{center}
		\includegraphics[width=0.85\columnwidth]{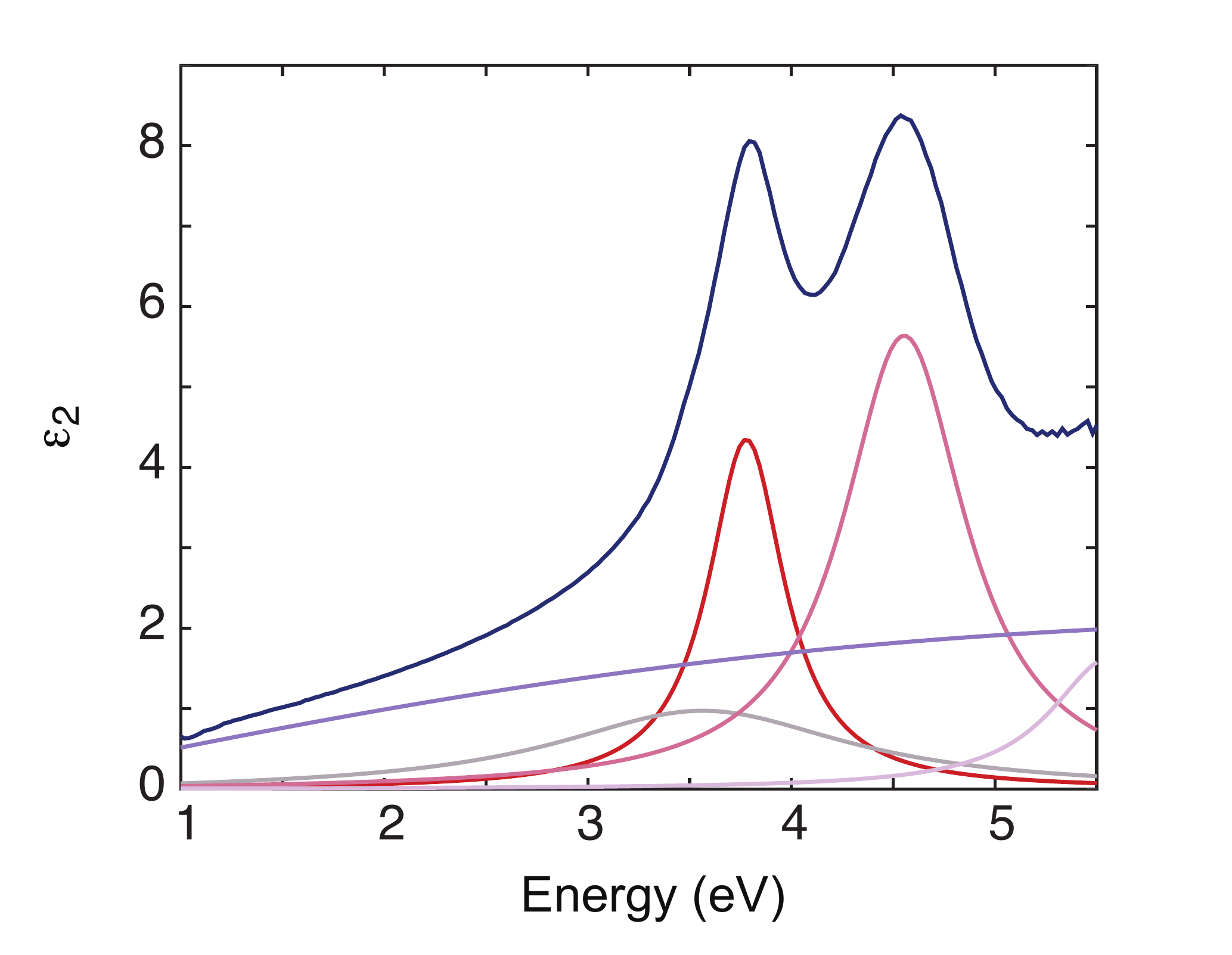}
		\caption{Decomposition of the $\epsilon_2(\omega)$ data (blue lines) into the separate Lorentz oscillators used in our model. The final fit is shown in Fig. S2(b).}
		\label{fig:FigS4}
	\end{center}
\end{figure}

Once this model of the steady-state optical response was established, we combined the steady-state data of $R(\omega)$ with the time-resolved reflectivity ($\Delta$R/R($\omega$,t)) measured in this work, following a well-known procedure in the context of ultrafast broadband optical spectroscopy \cite{mansart2012evidence, novelli2012ultrafast, palmieri}. We remark that the static $R(\omega)$ was measured in the broad spectral range 1.0-5.5~eV, while $\Delta R/R(\omega$,t) was monitored by our ultrafast experiment in the 3.6-4.4 eV region. Combining these two quantities allowed us to obtain the momentary reflectivity $R(\omega,t)$ in the range covered by the pump-probe experiment by multiplying ($\Delta R/R(\omega,t)$ + 1) at a fixed time delay $t$ by $R(\omega)$ itself. These are the raw data shown in Fig. 3(b) of the main text. Finally, to determine the momentary absorption $\alpha(\omega,t)$, we iterated the Lorentz fit at all measured time delays, using as starting parameters those describing the steady-state $R(\omega)$ of Fig. 1(b) and letting only the Lorentz oscillator representing the bound exciton free to vary. This was sufficient to reproduce the spectra at all time delays. The choice of letting only the bound exciton free to vary is justified by the fact that the pump excitation induces a very small modification of the reflectivity, depleting the optical spectral weight only around this exciton. The depleted spectral weight is eventually transferred to very low energies (i.e. mostly the terahertz range) in the form of free-carrier absorption \cite{matsuzaki2014photocarrier}. This procedure allowed us to retrieve all the optical quantities of interest using the standard electrodynamical formulas \cite{dressel2002electrodynamics}, among which the $\alpha$($\omega$,t) spectrum presented in Fig. 3(c), as well as the evolution of the bound exciton parameters (Fig. 3(d-f)).\\

\section{VI. High time resolution data}

We also performed high-precision measurements of $\Delta R/R(\omega,t)$ in a narrower spectral range around the exciton peak (3.72-4.20 eV) with a time resolution of 80~fs. The pump fluence is the same as that used in the pump-probe experiment with 700 fs time resolution. Figure~S5(a) shows normalized temporal traces selected at representative probe photon energies. Although the response was measured up to 20 ps, here we just display the first ps of dynamics. We observe a resolution-limited rise of the response in the low-energy region of the spectrum, followed by a fast relaxation component that is prominent around 3.95 eV. Figure S5(b) displays the momentary reflectivity $R(\omega,t)$ before photoexcitation (blue curve) and at a time delay of 120 fs (red curve). At 120~fs, we observe the persistence of the bound exciton feature, which becomes broader due to the larger long-range Coulomb screening caused by the photoexcited uncorrelated e-h pairs (note that the wiggles that are visible on the signal are due to noise). The fact that the exciton broadens implies a substantial modification of the exciton coherence lifetime, but not of the exciton $E_B$. As such, the estimated photoexcitation density represents a lower bound to the nominal $n_M$ in anatase TiO$_2$.

\begin{figure}[t]
	\begin{center}
		\includegraphics[width=0.7\columnwidth]{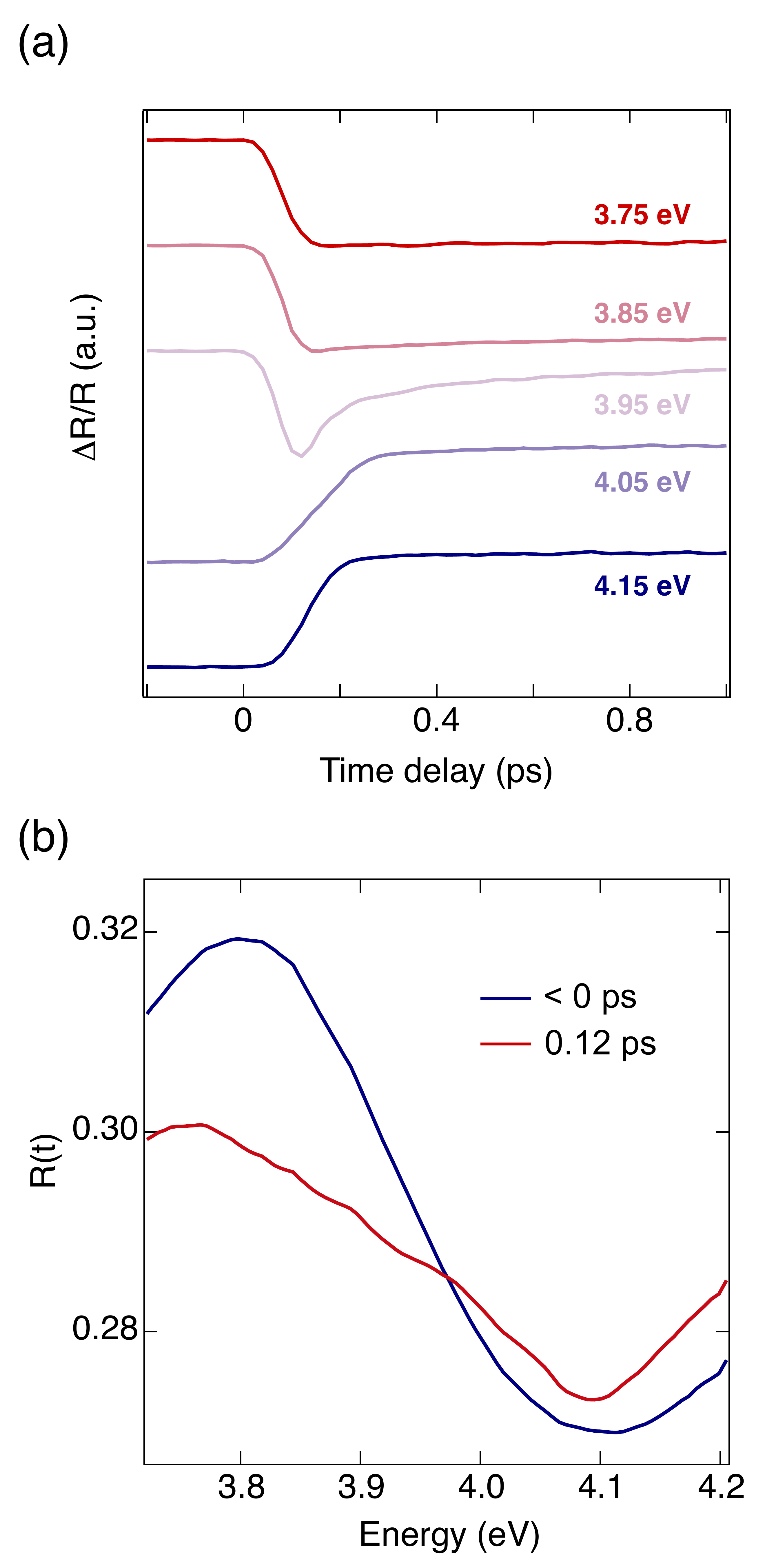}
		\caption{(a)High-time resolution $\Delta$R/R temporal traces selected at representative photon energies, as indicated in the labels. For clarity, the traces are normalized, vertically shifted, and displayed in the first ps of response. (b) Comparison between the momentary reflectivity $R(\omega,t)$ before and after photoexcitation.}
		\label{fig:FigS5}
	\end{center}
\end{figure}

\begin{table*}[tb]
	\begin{center}
		
		\begin{tabular}{|c|c|c|c|c|}
			\hline
			\textbf{Material} & \textbf{$E_B$ (meV)} & \textbf{$n_M$ (cm$^{-3}$)} & \textbf{Temperature (K)} & \textbf{Reference}\\
			\hline
			
			Anatase TiO$_2$ & $>$ 150 & $>$ 5 $\times$ 10$^{19}$	&	295 K &	This work\\

			ZnO & 60	& 0.8-6.4 $\times$ 10$^{18}$	&	295 K &	\cite{klingshirn2007room,versteegh2011ultrafast,schleife2011optical}\\
			
			CH$_3$NH$_3$PbBr$_3$  & 70 & 8 $\times$ 10$^{17}$	&	295 K & \cite{palmieri}\\
			
			GaAs & 4.2	& 1.2-1.8 $\times$ 10$^{16}$	&	49 K & \cite{amo2006interplay}\\
			
			GaN & 26 & 1 $\times$ 10$^{18}$	&	10 K & \cite{choi2001ultrafast}\\
			
			Diamond & 80	& 3 $\times$ 10$^{18}$	&	295 K &	\cite{nagai2003phase}\\
			
			Cu$_2$O & 150	& 3 $\times$ 10$^{18}$	&	10 K & \cite{manzke2010mott,heckotter2018rydberg}\\
			
			Si & 15	& 7 $\times$ 10$^{17}$	&	295 K & \cite{altermatt1997assessment,hull1999properties,green2013improved}\\
			
			Ge & 4.2	& 1 $\times$ 10$^{16}$	&	8 K & \cite{sekiguchi2015excitonic}\\
			
			ZnTe & 12.7	& 3 $\times$ 10$^{17}$	&	20 K & \cite{majumder1995carrier,yu2001photoluminescence}\\
			
			ZnS & 36-40	& 1.9 $\times$ 10$^{17}$	&	2 K & \cite{manar1997characteristic}\\
			
			ZnSe & 19-20	& 0.1-5 $\times$ 10$^{17}$	&	2-295 K & \cite{manar1997characteristic, semkat2009ionization}\\
			
			CdS & 28	& 2.6-50 $\times$ 10$^{17}$	&	2-295 K & \cite{henneberger1988exciton,fricke1994recombination}\\
			
			CdSe & 15	& 1 $\times$ 10$^{17}$	&	80 K & \cite{fujimoto1984femtosecond,maksimov2006luminescent}\\
			
			CdTe & 24.8	& 3 $\times$ 10$^{18}$	&	295 K & \cite{karazhanov2000effect}\\
			\hline
			
		\end{tabular}
		\caption{Exciton Mott densities in three-dimensional band semiconductors.}
		
	\end{center}
\end{table*}

\section{VII. Relevance of the high Mott density in anatase TiO$_2$}

In this Section, we describe the importance of our results in relation to fundamental and applied research.

\noindent Fundamentally, a large $n_M$ is a prerequisite for the creation of room temperature exciton-polaritons when the system is placed inside a microcavity. This has been explored in the case of Wannier excitons of low-dimensional semiconductor nanostructures \cite{deng2002condensation} and in the context of Frenkel excitons in molecular systems \cite{plumhof2014room,daskalakis2014nonlinear}. In anatase TiO$_2$, the excitons are characterized by large oscillator strength, small dephasing due to population relaxation into e-h pairs, and high $n_M$. These aspects together would allow the realization of stable exciton-polaritons in high-quality thin films of TiO$_2$ embedded in optical cavities and open the avenue to the study of possible condensation phenomena and nonlinear polariton-polariton interactions.

Technologically, a high $n_M$ ensures that the excitons are very stable quasiparticles even in the presence of the large carrier densities involved in many applications. One of these applications is photocatalysis, in which anatase TiO$_2$ represents one of the most used platforms at room temperature \cite{ref:fujishima}. It has been proposed that the excitons of anatase TiO$_2$ may allow for an efficient transfer of energy to the reaction centers at the (001) surfaces of conventionally-used TiO$_2$ nanoparticles \cite{varsano2017role}. The presence of a large $n_M$ ensures that the exciton population survives even in the presence of many thermally-activated carriers (owing to defects and impurities) in these highly-defective nanoparticles. Another application in which anatase TiO$_2$ is widely used is that of transparent conducting substrates. The conductivity of these substrates stems from the transport of thermally-activated electrons (owing to the presence of donor states, such as oxygen vacancies or Nb/Ta substitutions) \cite{furubayashi2005transparent,hitosugi2005ta}. Revealing that TiO$_2$ has such a high $n_M$ can also guide the rational design of the carrier density needed to optimize the transparency window of these substrates (by choosing densities above $n_M$, at which the bound exciton peak disappears and the material is more transparent). 

\section{VII. Comparison with other materials}

In this Section, we compare the results obtained on anatase TiO$_2$ with those of band insulators (i.e. devoid of strong electron-electron correlations) known in the literature. To establish an accurate comparison, we consider bulk solids characterized by a three-dimensional electronic structure and whose exciton states are electric dipole-allowed and built up from the mixing of single-particle transitions between the valence and the conduction bands. Hence, we do not include quasi-two dimensional materials ($e.g.$, transition metal dichalcogenides), confined nanostructures, as well as bulk solids with Frenkel excitons arising from localized states ($e.g.$, $d-d$ transitions in transition metal oxides with partially-filled $d$ orbitals). The complete list is shown in Table S1. Some of the materials considered here host excitons with low values of $E_B$ and therefore the estimate of $n_M$ had to be performed at low temperature. In contrast, other solids have strongly bound excitons and $n_M$ is evaluated at room temperature. We observe that anatase TiO$_2$ has the largest $n_M$, i.e. at least one order of magnitude higher than that exhibited by other band semiconductors. As indicated in the main text, we ascribe this phenomenon to the interplay between the large $E_B$ and the small impact of bandgap renormalization when an excess carrier density is injected in the solid (either through chemical- or photo-doping). 


\providecommand{\noopsort}[1]{}\providecommand{\singleletter}[1]{#1}

\end{document}